# Semileptonic form factors of heavy-light mesons from lattice QCD

Rajan Gupta, Tanmoy Bhattacharya and David Daniel

*T-8, MS-B285, Los Alamos National Laboratory, Los Alamos, NM 87545*

The form factors for the semileptonic decays of heavy-light pseudoscalar mesons of the type $D \to Ke\nu$ are studied in quenched lattice QCD at $\beta = 6.0$ using Wilson fermions. We explore new numerical techniques for improving the signal and study $O(a)$ corrections using three different lattice transcriptions of the vector current. We present a detailed discussion of the relation of these lattice currents to the continuum vector current and show that the disagreement between the previous results is to a large extent due to the value of $Z_V$ used in the calculations. We also present results for the decay constants of light-light, heavy-light and heavy-heavy mesons.

## 1. Introduction

The physics of mesons containing one heavy valence quark ($c$, $b$) and one light valence quark ($u$, $d$, $s$) is of considerable interest at present. Experimental investigations of these systems may provide the most accurate determination of the parameters of the standard model responsible for flavor mixing and $CP$ violation (the CKM matrix).* This can only be achieved, however, if we are able to obtain a quantitative understanding of the influence of the strong interaction on their structure and decays.

Numerical simulations of lattice QCD provide a solution to this problem, and there has been considerable activity in the field in recent years. Two groups have presented results for the form factors for the semi-leptonic decays of heavy-light pseudoscalar and vector mesons, namely Bernard *et al.* [2] [3] [4], and the Rome-Southampton group [5] [6] [7]. These results have large statistical errors, and, in certain instances, are in conflict. To resolve these discrepancies and check for systematic errors, further studies using different methodologies are necessary, and this is what we undertake in this paper.

In this paper we present data on the pseudoscalar and vector meson decay constants for light-light, heavy-light and heavy-heavy systems. We then move on to the main topic of this paper semileptonic form factors for pseudoscalar meson decays such as $D \to K e \nu$. The structure of this paper is the following. We review the phenomenology of semi-leptonic form-factors in Section 2. In Section 3 we discuss the renormalization of lattice operators and give details of the lattice setup in Section 4. Results for the decay constants are presented in Section 5 along with those for the vector current renormalization $Z_V$. The analysis of the form-factors is given in Section 6 along with a discussion of the quality of the signal and a comparison with previous data is given in Section 7. Finally, we state our conclusions in Section 8.

## 2. Phenomenological Background

We consider the case, $D \to X l \nu$, where $X$ has flavor content $\overline{u}s$ ($K$ or $K^*$). In the one $W$ exchange approximation the amplitude is

$$\begin{aligned}\langle X^- l^+ \nu | \, H_W \, | D^0 \rangle &= \frac{G_F}{\sqrt{2}} \int d^4x \, \langle X^- l^+ \nu | \, (V-A)^\dagger_\mu (V-A)_\mu \, | D^0 \rangle, \\ &= \frac{G_F}{\sqrt{2}} V_{sc} \, \overline{v}(l) \gamma_\mu (1-\gamma_5) u(\nu) \langle X^- | \, \overline{s} \gamma_\mu (1-\gamma_5) c \, | D^0 \rangle,\end{aligned} \tag{2.1}$$

* A recent overview of the experimental status of charmed meson physics is given in ref. [1].

where $G_F$ is the Fermi constant, $V_{cs}$ is the $c \to s$ CKM matrix element. This process is particularly simple because the hadronic and leptonic currents factorize. The leptonic part of the decay can be calculated accurately using perturbation theory, while to take into account non-perturbative contributions to the hadronic part

$$H_\mu = \langle X | \bar{s}\gamma_\mu(1-\gamma_5)c | D\rangle \tag{2.2}$$

one resorts to lattice QCD. In this paper we present results for the simpler of the two cases, *i.e.* $D^0 \to K^- e^+ \nu$.

The matrix element $H_\mu$ can be parameterized in terms of two form factors:

$$\left\langle K^-(p_K)\right| \bar{s}\gamma_\mu(1-\gamma_5)c \left| D^0(p_D)\right\rangle = p_\mu f_+(Q^2) + q_\mu f_-(Q^2), \tag{2.3}$$

where $p = (p_D + p_K)$ and $q = (p_D - p_K)$ is the momentum carried away by the leptons, and $Q^2 = -q^2$ (which is always positive). We use the Euclidean notation $p = (\vec{p}, iE)$ so that $p^2 = \vec{p}^{\,2} - E^2$. An alternative parameterization is

$$\begin{aligned} &\left\langle K^-(p_K)\right| \bar{s}\gamma_\mu(1-\gamma_5)c \left| D^0(p_D)\right\rangle \\ &\qquad = \left(p_\mu - \frac{m_D^2 - m_K^2}{Q^2} q_\mu\right) f_+(Q^2) + \frac{m_D^2 - m_K^2}{Q^2} q_\mu f_0(Q^2), \end{aligned} \tag{2.4}$$

where

$$f_0(Q^2) = f_+(Q^2) + \frac{Q^2}{m_D^2 - m_K^2} f_-(Q^2). \tag{2.5}$$

In the center of mass coordinate system for the lepton pair, *i.e.* $\vec{q} = 0$ or equivalently $\vec{p}_K = \vec{p}_D$, one has

$$\begin{aligned} \left\langle K^-(p_K)\right| \bar{s}\vec{\gamma}c \left| D^0(p_D)\right\rangle &= 2\vec{p}_D f_+(Q^2), \\ \left\langle K^-(p_K)\right| \bar{s}\gamma_4 c \left| D^0(p_D)\right\rangle &= \frac{m_D^2 - m_K^2}{\sqrt{Q^2}} f_0(Q^2). \end{aligned} \tag{2.6}$$

Thus, the form factor $f_+(Q^2)$ is associated with the exchange of a vector particle, while $f_0(Q^2)$ is associated with a scalar exchange. It is common to assume nearest pole dominance and make the hypothesis

$$f_+(Q^2) = \frac{f_+(0)}{1 - Q^2/m_{1^-}^2}, \qquad f_0(Q^2) = \frac{f_0(0)}{1 - Q^2/m_{0^+}^2}, \tag{2.7}$$

where $m_{J^P}$ is the mass of the lightest resonance with the right quantum numbers to mediate the transition; $D_s^+(1969)$ or $D_s^{*+}(2110)$ in the pseudoscalar or vector channels respectively.

The goal of the lattice calculations is to determine the normalizations $f_+(0)$ and $f_0(0)$ and map out the $Q^2$ dependence.

In the limit of vanishing lepton masses, the vector channel dominates and one can write the the differential decay rate as

$$\begin{aligned} d\Gamma(Q^2) &= \frac{G_F^2|V_{cs}|^2}{192\pi^3 m_D^3} dQ^2 \lambda(Q^2)^{3/2} |f_+(Q^2)|^2, \\ \lambda(Q^2) &= (m_D^2 + m_K^2 - Q^2)^2 - 4m_D^2 m_K^2. \end{aligned} \tag{2.8}$$

To integrate this, the functional form of $f_+$ must be known. Assuming vector meson dominance, numerical integration gives

$$\Gamma(D^0 \to K^- e^+ \nu) = 1.53 |V_{cs}|^2 |f_+(0)|^2 \times 10^{-11} \mathrm{sec}^{-1}. \tag{2.9}$$

Using Eqn. (2.9) one can extract $V_{cs}$ once $\Gamma(D^0 \to K^- e^+ \nu)$ has been measured and $f_+$ calculated using lattice QCD. In this case $D^0 \to K^- e^+ \nu$, however, $|V_{cs}| = 0.975$ is known very accurately, so one extracts $|f_+(0)| \approx 0.75$. The quantity $f_0(0)$ has not been determined.

The present study, whose goal is to investigate different numerical techniques in order to improve the signal to noise ratio, shows that first principles calculation of form-factors can be carried out reliably with today's massively parallel computers.

## 3. Renormalization of lattice operators

Lattice transcriptions of continuum operators like the vector and axial currents suffer from $O(a)$ corrections. To get a handle on these $O(a)$ effects in the study of form factors, we use three transcriptions for the vector current

$$\begin{aligned} V_\mu^{\mathrm{local}}(x) &= \overline{q}_1(x)\gamma_\mu q_2(x), \\ V_\mu^{\mathrm{ext.}}(x) &= \frac{1}{2}\left(\overline{q}_1(x)\gamma_\mu U_\mu(x) q_2(x+a\mu) + \overline{q}_1(x+a\mu)\gamma_\mu U_\mu(x)^\dagger q_2(x)\right), \\ V_\mu^{\mathrm{cons.}}(x) &= \frac{1}{2}\left(\overline{q}_1(x)(\gamma_\mu - 1) U_\mu(x) q_2(x+a\mu) + \overline{q}_1(x+a\mu)(\gamma_\mu + 1) U_\mu(x)^\dagger q_2(x)\right). \end{aligned} \tag{3.1}$$

We remind the reader that $V_\mu^{\mathrm{cons.}}(x)$ is conserved only for degenerate quarks. In the calculation of the pseudoscalar decay constant we use only the local axial current $\overline{q}_1(x)\gamma_5\gamma_\mu q_2(x)$. In these currents the quarks $q_1$ and $q_2$ may both be light, or one heavy and one light, or both heavy. The first step in relating these lattice currents to their continuum counterparts is to calculate the normalization of lattice fields.

Lepage and Mackenzie have given a prescription for relating lattice quantities to their continuum counterparts [8]. They show how to handle the large tadpole contributions to operators and as an example show that the renormalization constant $Z_V$ for the local vector current agrees very well with non-perturbative measurements even for $ma \approx 1$ [9]. Since the reliablity of calculations of matrix elements depends crucially on accurate determination of these parameters, we present an explicit 1-loop analysis and reproduce the results of Lepage and Mackenzie.

The overall philosophy of the $O(\alpha_s)$ improvement scheme is that in perturbation theory tadpole contributions are large when quantities are expressed in terms of the bare parameters. The three basic ways in which tadpoles contribute to Feynman diagrams are shown in Fig. 1. These can and should be absorbed into the renormalized quantities, so that the resulting perturbation theory is well behaved. The goal is to absorb the correction to the fermion line in the definition of the renormalized mass, to the gluon line in the definition of the renormalized coupling, and whenever possible to cancel the tadpole contributions in operators by combining them with the renormalized quantities. We now show how this works to order $\alpha_s$ for the different currents we use in this paper.

### *3.1. Field renormalization*

We want to calculate $Z_\psi$, the relative normalization of lattice fields, defined as

$$\psi_{\text{continuum}} = \sqrt{Z_\psi}\ \psi_{\text{lattice}}. \tag{3.2}$$

On dimensional grounds, $Z$ can only be a function of $r$, $\alpha_s$ and $ma$ for Wilson fermions, where $ma$ is the quark mass in lattice units. The inverse of the Wilson propagator is (we set $r = 1$ to simplify the discussion)

$$(2\kappa\mathcal{P})^{-1} = \frac{1}{a}\sum_\mu i\gamma_\mu \sin k_\mu a + \frac{2}{a}\sum_\mu \sin^2\frac{k_\mu a}{2} + m + \frac{1}{a}\alpha_s f(k_\mu a, ma) + O(\alpha_s^2) \tag{3.3}$$

where the factor of $1/a$ has been extracted to make it dimensionless. The 1-loop contribution $f$ has the structure $f = \Sigma_0 + \Sigma_1 \not{k}a + \Sigma_2 ma$ and it contains the tadpole contribution $(i\not{k}a - 4)K$ [10] [11]. The $\Sigma_i$ and $K$ are numerical constants. In this equation $ma$ is defined to be the bare value $1/2\kappa - 4$. We shall keep track of terms of $O(\alpha_s)$ and $O(ma)$, but neglect $O(\alpha_s^2)$ and $O(\alpha_s ma)$ terms. Thus, the results are strictly speaking valid only for $\alpha_s ma < \alpha_s$, though in practice the range of validity has to be determined non-perturbatively.

We now set $\vec{k} = 0$ and express quantities in terms of $k_0 = ik_4$. For brevity of notation we write $f(-ik_0, ma)$ as $f(k_0, ma)$. Then the forward moving part of the inverse propagator is

$$(2\kappa\mathcal{P})^{-1} a = -\exp k_0 a + 1 + ma + \alpha_s f(k_0 a, ma). \tag{3.4}$$

The renormalized mass is defined by the location of the pole in the propagator, which is given by the relation

$$\exp k_0 a = 1 + ma + \alpha_s f(k_0 a, ma). \tag{3.5}$$

The zero mass limit is determined from the condition ($m_c \equiv 1/2\kappa_c - 4$)

$$m_c a + \alpha_s f(0, m_c a) = 0 \tag{3.6}$$

which is used to define the value of $\kappa_c$

$$\begin{aligned} \frac{1}{2\kappa_c} &= 4 - \alpha_s f(0, m_c a) \\ 8\kappa_c &= 1 + \frac{1}{4}\alpha_s f(0, m_c a) \end{aligned} \tag{3.7}$$

Close to the chiral limit $ma$ is of order $\alpha_s$ as can be seen from Eq. (3.6). In order to calculate the wave-function renormalization one needs to evaluate the residue at the pole. This is

$$\begin{aligned} \exp k_0 a - \alpha_s \frac{\partial f(k_0 a, 0)}{\partial k_0 a} &= 1 + ma + \alpha_s f(k_0 a, 0) - \alpha_s \frac{\partial f(k_0 a, 0)}{\partial k_0 a} \\ &\approx [1 + ma + \alpha_s f(k_0 a, 0)][1 - \alpha_s \frac{\partial f(k_0 a, 0)}{\partial k_0 a}] \end{aligned} \tag{3.8}$$

where we have set $ma = 0$ in terms proportional to $\alpha_s$ as these are $O(\alpha_s ma)$. Note that all the terms proportional to $\alpha_s$ have contributions from tadpole diagrams. In lattice calculations it has been traditional to factor the residue into two terms, the first term is a lattice effect and defined to be the relative renormalization of the lattice to continuum field $\psi$ and the second is the square of the wave-function renormalization that has to be included for each external fermion leg in the operator. The seemingly bad behavior of perturbation theory arises if these two terms are truncated at different orders of $\alpha_s$. We consider it useful to give explicit details of how the tadpole contributions are distributed and absorbed into the renormalized quantities so that the resulting perturbation expansion is well-behaved.

The first step in the reorganization is to renormalize the bare quark mass and express it in terms of $\kappa_c$. Using Eq. (3.7) we get

$$\begin{aligned} 1 + ma + \alpha_s f(k_0 a, 0) &= \frac{1}{2\kappa} - 3 + \alpha_s f(k_0 a, 0) \\ &= \frac{4\kappa_c}{\kappa}\left[1 - \frac{3\kappa}{4\kappa_c}\right] + \alpha_s\left(f(k_0 a, 0) - \frac{f(0, m_c a)}{8\kappa}\right). \end{aligned} \tag{3.9}$$

Now the term proportional to $\alpha_s$ is of order $\alpha_s ma$, and can therefore be dropped. This rearrangement gives the final result for the residue of the propagator $\mathcal{P}$ (note that we have to multiply Eq. (3.8) by $2\kappa$), *i.e.*

$$8\kappa_c\left[1 - \frac{3\kappa}{4\kappa_c}\right]\left[1 - \alpha_s\frac{\partial}{\partial k_0 a} f(k_0 a, 0)\right], \tag{3.10}$$

which is to be evaluated at the pole. The tadpole contributions are distributed as follows. As shown in Eq.(3.7) the factor $8\kappa_c$ absorbs 1/4 of the tadpole term that is not proportional to $\not{k}$ . This is equal and opposite in sign to the tadpole term proportional to $\not{k}$ that is included in the standard wave-function renormalization $\big(1 - \alpha_s \partial f(k_0 a, 0)/\partial k_0 a\big)$. Therefore these two pieces always cancel. We reiterate this observation in slightly different words to emphasize the point. The tadpole contribution to $8\kappa_c$ in the perturbative expansion

$$8\kappa_c = 1 + 1.364\alpha_V + \ldots \tag{3.11}$$

is equal and opposite to that from wave-function renormalization of the external legs. This is true for any operator $O$. Thus the combination $8\kappa_c Z_O$ will be better behaved at $O(\alpha_s)$. The remaining tadpole part is included in $\sqrt{1 - 3\kappa_i/4\kappa_c}$. On using the measured value for $\kappa_c$ this factor becomes independent of perturbation theory and is therefore well behaved. This completes our discussion of tadpoles on the fermion line. A way to include tadpoles on gluon lines will be discussed later when we describe the Lepage-Mackenzie scheme.

Using this scheme we can now write down the normalizations for the vector current. The 1-loop calculations of the local vector and axial currents have been done by Martinelli and Zhang [10] and we have checked their calculations. The local vector current contains a tadpole contribution from the wave-function renormalization, so we combine it with the perturbative expansion for $8\kappa_c$ to get

$$V_\mu\Big|_{cont} \equiv \overline{q}_1(x)\gamma_\mu q_2(x)\Big|_{cont} = \sqrt{1 - \frac{3\kappa_1}{4\kappa_c}}\sqrt{1 - \frac{3\kappa_2}{4\kappa_c}}(1 - 0.82\alpha_V)\overline{q}_1(x)\gamma_\mu q_2(x)\Big|_L . \tag{3.12}$$

Here $\alpha_V$ is the renormalized coupling that is defined later. To get the normalization for extended currents one has also to include the $O(ma)$ terms in the vertex. For the conserved current the vertex at tree level is modified to $e^{-k_0 a}\gamma_\mu$. The extra factor $e^{-k_0 a}$ has been obtained by making the same simplifications as in Eq. (3.4). This factor exactly cancels the $[1+ma+\alpha_s f(k_0 a,0)]$ part of Eq. (3.8). The remainder $(1-\Sigma_1)$ is equal and opposite in sign to the 1-loop corrections to the vertex. Thus the renormalization factor for the "conserved" current is unchanged from the tree level definition†

$$V_\mu\Big|_{cont} = \sqrt{2\kappa_1}\sqrt{2\kappa_2}V_\mu^{cons.}\Big|_L . \tag{3.13}$$

The above result also applies to the case when the current is flavor changing but its matrix elements are taken in the forward direction. On the other hand the vertex for the extended 1-link current is $\gamma_\mu \cos k_\mu a$, which reduces to $\gamma_\mu$ at the order to which we are working. The two tadpole contributions, one from the wave-function renormalization $1-\Sigma_1$ and the other from the expansion of the link in the operator are again equal and opposite. Since these two cancel one does not need to combine the perturbation result for $Z_V^{ext.}$ with that for $8\kappa_c$. Thus we use

$$V_\mu\Big|_{cont} = 8\kappa_c\sqrt{1-\frac{3\kappa_1}{4\kappa_c}}\sqrt{1-\frac{3\kappa_2}{4\kappa_c}}(1-1.038\alpha_V)V_\mu^{ext.}\Big|_L . \tag{3.14}$$

It should be emphasized that our analysis does not specify whether it is better to use the perturbative or non-perturbative value for $8\kappa_c$ in cases where the tadpoles in the perturbative part cancel. The difference between using the perturbative and non-perturbative result is $\approx 6\%$ at $\beta = 6.0$, *i.e.* $8\kappa_c^{pert.} = 1.185$ versus the non-perturbative value of 1.256. This difference is a measure of the residual uncertainties in the renormalization of the currents.

Lastly, the vertex for the local axial current gets a tadpole contribution only from $(1-\Sigma_1)$. This can be cancelled by combining it with the perturbative expansion for $8\kappa_c$. Then

$$A_\mu\Big|_{cont} = \sqrt{1-\frac{3\kappa_1}{4\kappa_c}}\sqrt{1-\frac{3\kappa_2}{4\kappa_c}}(1-0.31\alpha_V)A_\mu\Big|_L . \tag{3.15}$$

Lepage and Mackenzie have given a mean-field prescription for determining the relation between lattice and continuum quantities [8]. The only difference in the final results for local operators in the mean-field approach is to replace the prefactor $8\kappa_c$ by

† We thank G. Martinelli for discussions on this point.

$1/U_0 \equiv 1/U_{plaq}^{1/4}$ where $U_{plaq}$ is the trace of the plaquette. However, within the mean-field framework $8\kappa_c U_0 = 1$ and deviations from unity are a measure of uncertainty in this approach. To give the reader a feel for how consistent these approaches are we give the numerical values at $\beta = 6.0$. They are $1/U_0 = 1.140$ and $8\kappa_c U_0 = 1.1$. The perturbative result for $1/U_0$ is $1 + 1.047\alpha_V = 1.165$.

To summarize the Lepage-Mackenzie prescription for removing the potentially large tadpole contributions in local fermionic operators at $O(\alpha_s)$ the cook-book recipe is

(a) For each quark of flavor $i$ in the operator use the normalization

$$\psi^i_{cont} = \sqrt{8\kappa_c}\sqrt{1 - \frac{3\kappa_i}{4\kappa_c}}\;\psi^i_L \tag{3.16}$$

between the lattice field and its continuum counterpart.

(b) Calculate the $\alpha_s$ corrections to the operator $O$, both on the lattice and in the continuum in say the $\overline{MS}$ scheme. In the result $Z_O$ replace $\alpha_{bare}$ by $\alpha_V$ where [8]

$$\alpha_V(\frac{\pi}{a}) = \frac{\alpha_{lat}}{\langle 1/3\mathrm{Tr}U_{plaq}\rangle}\big(1 + 0.513\alpha_V + O(\alpha_V^2)\big). \tag{3.17}$$

This replacement removes the large tadpole contributions to gluon lines that would otherwise show up at $O(\alpha_s^2)$ and make the series look badly behaved.

(c) Determine the scale $q$ at which $\alpha_V$ is defined (one prescription for doing this is given in Ref. [8]) and scale $\alpha_V(\pi/a)$ to $\alpha_V(q)$ using 2-loop running. With these three things in hand one has the final relation $Z_O$ between the lattice and continuum operator to $O(\alpha_s)$. In the rest of the paper we shall choose $\alpha_V = g_R^2/4\pi$ as the renormalized coupling at scale $q = 1.4\pi/a$, where $g_R^2 = 1.7g_{bare}^2$ at $\beta = 6.0$.

(d) To extend the analysis to non-local operators one has to include the $O(ma)$ tree level corrections to the operator. These have to be combined with the renormalization of the field $\psi$ at the stage of Eq.(3.8). This can change the relation between the lattice and continuum field $\psi$ from that given in Eq. (3.16) as we have shown by the examples of the two 1-link vector currents.

The lesson from our analysis is that one cannot apply the mean-field prescription to calculate the renormalization factors for extended currents by simply counting the number of $\psi$ fields and links in the operator. Instead one has to work self-consistently to a given order in $\alpha_s$ and include $O(ma)$ and $O(pa)$ terms present in the Taylor expansion of the operator.

We shall present results using the improved normalizations given above and make comparisons with the naive scheme (without the improvement incorporating tadpoles on fermion lines), *i.e.* $\psi^i_{cont} = \sqrt{2\kappa^i}\ \psi^i_L$. In both cases, however, we will use the boosted $g^2$ in $\alpha_V$ to sum up the tadpole contribution to the gluon line. To allow the reader to evaluate the difference between the two normalizations we give the normalization factor for both schemes and their ratio in Table 1. As the numbers show, the difference due to the choice of the normalization of the currents grows rapidly with the quark mass.

Our perturbative analysis is only valid close to the chiral limit and takes into account $O(ma)$ effects only for on-shell quarks and for processes with no momentum transfer, *i.e.* it is not $O(a)$ improved. These conditions are not realized in the semi-leptonic decays of heavy quarks. Since the normalization factors have a large uncertainty especially at large quark mass, it is necessary to check the perturbative renormalization factors by non-perturbative methods where possible. El Khadra *et al.* [9] have measured $Z_V^{local}$ by calculating the matrix element $\langle J/\psi|\, V_4^{local}\, |J/\psi\rangle$ for heavy quarks ($m_q a \sim 1$) and find that Eq. (3.12) works incredibly well. This encouraging result needs to be verified for a variety of different matrix elements and we discuss this issue further in Section 5.

### *3.2. Extension to Sheikholeslami-Wohlert Action*

The above analysis can be easily extended to improved actions. The case of the $O(a)$ improved action proposed by Sheikholeslami and Wohlert [12] is particularly simple. This is because the tadpole contributions to the propagator with the SW action are the same as those for the Wilson action. Secondly, the perturbative expansion for $\kappa_c$ is unchanged. Thus the expression (3.10) is still valid except that the extra fermion-gluon vertex changes the function $f$ and consequently the precise form of $1 - \Sigma_1$. Similarly, the finite parts of the 1-loop result for the vertex are also different. These corrections have been calculated in Ref. [13] and all one needs to do is extract the appropriate perturbative corrections to a given operator from it. The rest of the renormalization factors are the same as for Wilson fermions and are given by the steps (a,c,d) described above.

The numerical data show that the improved normalizations work better for the SW action as a consequence of $O(a)$ improvement. For example, at $\beta = 6.0$ the non-perturbative value for the chiral limit has been measured to be $\kappa_c = 0.14556(6)$ [14]. Thus $8\kappa_c = 1.1645$ agrees very well with the perturbative result 1.185 and $8\kappa_c U_0 = 1.02$.

## 4. Details of the lattice setup

### *4.1. Lattice parameters*

Our statistical sample consists of 35 lattices of size $16^3 \times 40$ at $\beta = 6.0$. This sets of lattices have been used previously for spectrum and weak matrix element analysis using both Wilson and staggered fermions [15] [16] [17], and the details of the lattice generation are given in [15].

In this study our main goal is to investigate different numerical techniques in order to improve the signal to noise ratio. For this purpose we use only one value of the heavy quark mass, $\kappa = 0.135$, and only two values of the light quark mass, $\kappa = 0.154$ and 0.155. Our results for meson masses expressed in lattice units are given in Table 2. Using $a^{-1} = 1.9$ GeV, these corresponds to a heavy meson of mass 1.54 and 1.59 GeV (about the mass of the physical charm quark) and to light-light pseudoscalar masses of roughly 690 MeV and 560 MeV. Our heavy-light pseudoscalar mesons therefore correspond most closely to the physical $D$ meson, with a somewhat massive light constituent, while the light-light mesons are analogous to the physical $K$. We will henceforth adopt this nomenclature.

### *4.2. Quark propagators*

The calculation of quark propagators is done on lattices doubled in the time direction, *i.e.* $16^3 \times 40 \to 16^3 \times 80$. We use periodic boundary conditions in all four directions. These propagators on doubled lattices are identical to a linear combination of propagators calculated with periodic ($P$) and antiperiodic ($A$) boundary conditions on the original $16^3 \times 40$ lattice. For the source on time slice 1, the forward moving solution (time slices 2–40) is $F = (P + A)/2$ while the backward moving solution (time slices 80–42) is $B = (P - A)/2$.

The details of our implementation of the "Wuppertal" smeared source method [18] are described in [16] and we have reused the light-quark propagators generated in earlier calculations. For a given light quark mass we have four types of quark propagator which we denote $G^l_{LS}(x;0)$, $G^h_{LS}(x;0)$, $G^l_{SS}(x;0)$ and $G^h_{SS}(x;0)$ for "light local-smeared", "heavy local-smeared", "light smeared-smeared" and "heavy smeared-smeared". The "smeared-smeared" quark propagators are obtained by applying the Wuppertal smearing procedure at each sink time slice of the "local-smeared" propagators.

For the calculation of 3-point functions we require a fifth type of quark propagator, namely a light-heavy propagator with the insertion of a zero 3-momentum smeared pseudoscalar source at some fixed time $t_0$. This propagator, which we denote $G^{hl}_{t_0}$, is calculated by doing a heavy quark inversion using a zero-momentum solution of the light propagator on a specified time-slice as the source. In terms of quark propagators without insertions we have the definition

$$G^{hl}_{t_0}(x;0) = \sum_{\vec{y}} G^h_{LS}(x;\vec{y},t_0)\gamma_5 G^l_{SS}(\vec{y},t_0;0). \tag{4.1}$$

Note that we perform "Wuppertal" smearing twice at $t_0$ in the construction of $G^{hl}$. This is needed to cancel the smearing factors when constructing ratios of 3-point to 2-point correlators. We choose $t_0 = 32$ in order to be far from the boundary at $t = 40$. Thus wrap-around effects in time direction are exponentially damped by at least 18 time slices because we use propagators calculated on doubled lattices.

Our method for extracting form-factors is similar to that used by the Rome-Southampton group [5] [6] [7] except that we have used smeared quark propagators which improve the signal and calculate the form-factors for three different currents. In Sections 6 and 7 we comment on the improvements over previous results and make a detailed comparison.

### *4.3. 2-point and 3-point correlators*

The large time behavior of a 2-point correlator at a given 3-momentum is

$$\begin{aligned} C_{12}(\vec{p},t) &= \sum_{\vec{x}} \exp(-i\vec{p}\cdot\vec{x})\langle \mathcal{O}_2(\vec{x},t)\mathcal{O}_1(0)^\dagger\rangle, \\ &\sim \frac{e^{-E_h(\vec{p})t}}{2E_h(\vec{p})}\langle 0|\,\mathcal{O}_2\,|h(\vec{p})\rangle\langle h(\vec{p})|\,\mathcal{O}_1^\dagger\,|0\rangle, \qquad t\to\infty \\ &= \frac{\sqrt{Z_1(\vec{p})Z_2(\vec{p})}}{2\sqrt{m_{12}^2+\vec{p}^{\,2}}}\exp\left(-t\sqrt{m_{12}^2+\vec{p}^{\,2}}\right). \end{aligned} \tag{4.2}$$

where $\mathcal{O}_1$ and $\mathcal{O}_2$ are interpolating operators for the hadronic state, $|h\rangle$ of mass $M_h$ ($E_h(\vec{p})^2 = \vec{p}^{\,2} + M^2$), and $Z_i$ is the amplitude for creating and annihilating the state. The subscript $i$ in $Z_i$ stands for both the flavor of the quarks and the type of smearing for the source and sink. We have collected data using both the local and smeared pseudoscalar density and axial vector current

$$\begin{aligned} P(x) &= \overline{q}_1(x)\gamma_5 q_2(x), \\ A_\mu(x) &= \overline{q}_1(x)\gamma_\mu\gamma_5 q_2(x), \end{aligned} \tag{4.3}$$

to interpolate for pseudoscalar mesons. The three point function giving the matrix element of current, $J_\mu$, between hadronic states is defined to be:

$$\begin{aligned} C^{ins}_{t_0}(\vec{p},t) &= \sum_{\vec{x},\vec{y}} \exp(-i\vec{p}\cdot\vec{x})\langle \mathcal{O}_2(\vec{y},t_0) J_\mu(\vec{x},t))\mathcal{O}_1(0)^\dagger\rangle, \\ &\sim \frac{e^{-E_{h_1}(\vec{p})(t_0-t)-M_{h_2}t}}{4M_{h_2}E_{h_1}(\vec{p})} \langle 0|\,\mathcal{O}_2\left|h_2(\vec{0})\right\rangle\left\langle h_2(\vec{0})\right| J_\mu\,|h_1(\vec{p})\rangle\langle h_1(\vec{p})|\,\mathcal{O}_1^\dagger\,|0\rangle, \end{aligned} \tag{4.4}$$

for $0 \ll t \ll t_0$.

In order to extract $H_\mu$ we construct the following two ratios of 3-point to 2-point correlation functions. For the case $\mathcal{O}_1 = \mathcal{O}_2 = P$ and $J_\mu = V_\mu^{local}$ in Eqs. (4.2) (4.4), and using the $LS$ heavy-light and light-light 2-point correlators, the ratio $\mathcal{R}_{LS}(\vec{p})$ is defined to be

$$\begin{aligned} \mathcal{R}_{LS}(\vec{p}) &\equiv \frac{C^{\rm ins}_{t_0}(\vec{p},t)}{C^{ll}_{LS}(\vec{p},t)C^{hl}_{LS}(\vec{0},t_0-t)} \\ &\sim \sqrt{Z^{ll}_{LS}(\vec{p})Z^{hl}_{LS}(\vec{0})}\,\langle K(p)|\,\bar{s}\gamma_\mu c\left|D(\vec{0})\right\rangle, \quad 0<t<t_0. \end{aligned} \tag{4.5}$$

Similarly, using the $SS$ heavy-light and light-light 2-point correlators gives

$$\begin{aligned} \mathcal{R}_{SS}(\vec{p}) &\equiv \frac{C^{\rm ins}_{t_0}(\vec{p},t)}{C^{ll}_{SS}(\vec{p},t)C^{hl}_{SS}(\vec{0},t_0-t)} \\ &\sim \sqrt{Z^{ll}_{SS}(\vec{p})Z^{hl}_{SS}(\vec{0})}\,\langle K(p)|\,\bar{s}\gamma_\mu c\left|D(\vec{0})\right\rangle, \quad 0<t<t_0. \end{aligned} \tag{4.6}$$

The meson decay amplitudes in Eqs. (4.5) and (4.6) are determined in separate fits. The results of these fits are combined at the time of jackknife analysis to extract $H_\mu$. This procedure is described in more detail in the next sub-section. Note that the two ways of calculating $H_\mu$ differ only in the 2-point correlators used to cancel the exponential dependence on $t$ and subsequently on which meson decay amplitudes have to be removed. Thus the difference between the results is a measure of the uncertainty in the fits. We show results for the two cases separately and for the average of the two as it reduces the fluctuations. The normalization of the currents is incorporated at the very end of the analysis as it is a common overall factor.

### *4.4. Fitting procedure*

To extract amplitudes and masses from the large time behavior of 2- point correlators we assume that at large $t$ only the lowest mass state dominates the correlation functions. To ensure this we first examine the effective mass plot for a plateau and then make a single mass fit selecting the range of the fit based on the following criteria: (a) $t_{\rm min}$ always lies in the plateau, (b) $t_{\rm max}$ is selected to be as large as possible consistent with a signal.

In most cases we find that the central value obtained from the fits is the same with and without using the full covariance matrix. In some cases we cannot use the full range of the plateau because the covariance matrix is close to being singular. In such cases the problem is not that we cannot invert the covariance matrix but that the result is very sensitive to the range of the fit (the central value can change by one or more standard deviations on the addition of a single point to the fit range) and/or some of the jackknife samples give grossly different results. Our tests using subsets of the data and by examining the covariance matrix show that this instability is a result of inadequate statistics. Some of the causes for this bad behavior are discussed in Ref. [19]. To overcome this problem we reduce the range of the fit until we get a stable result and compare it with that obtained without the full covariance matrix to check for consistency.

All error estimates are obtained using a single-elimination jackknife procedure. Previous analysis [20] leads us to believe that the lattices used are sufficiently decorrelated for this method to be adequate.

In order to extract the matrix element $H_\mu$ we make two separate fits for each jackknife sample to (1) the ratios of correlators defined in Eqs. (4.5) and (4.6) assuming that the result is constant over times slices on which the lowest state saturates the 2-point correlation functions; (2) the 2-point correlators needed to extract the extra decay amplitudes that have to be removed from $\mathcal{R}_{LS}$ and $\mathcal{R}_{SS}$. The results of the fits are combined for each jackknife sample and the final value and error on it is calculated by the jackknife method. This process is explained in more detail in Sections 5 and 6.

## 5. Meson decay constants

### *5.1. Pseudoscalar mesons*

The definition of the pseudo-scalar decay constant on the lattice is [21]

$$f_\pi = \frac{Z_\psi^2 Z_A \langle 0| \, A_4^{\rm local} \, |\pi(\vec{p})\rangle}{E_\pi(\vec{p})}, \qquad (5.1)$$

where $Z_\psi^2 Z_A$ is the axial current renormalization, and $f_\pi = 132$ MeV in our convention. The details of the four different methods we use to extract the decay amplitude are given in Ref. [16], so we only present the results here. All four methods give consistent estimates of $f_\pi$ and we quote their mean in Table 3 for three different choices of $Z_\psi^2 Z_A$.

Since we combine results from different correlators in order to extract $f_\pi$ we select the fit range based on the following criteria: (1) goodness of the fit, (2) presence of a plateau with a similar mass estimate from each of the individual correlators. To monitor for the possibility that the covariance matrix is not well determined we make fits with and without the full covariance matrix. In cases where the results with the covarience matrix deviate from the data we decrease the range of the fit to see if the results stabilize, otherwise we quote the central value from the uncorrelated fits and give the larger of the two error estimates.

Results for $f_\pi$ at $\vec{p} = (0, 0, 2\pi/16)$ are also given in Table 3 and are consistent with $\vec{p} = 0$ data. The quality of the signal is good enough to extract reliable numbers even though the plateaus in the effective mass are much shorter than those for the $\vec{p} = (0, 0, 0)$ case. At higher momenta the errors get much larger and the determination of $f_\pi$ is not reliable.

Some of the data in Table 3 have been published previously in Ref. [16]. The old numbers may differ from those given in Table 3 under label C because we reanalyzed all the data and the subjective choice of the range of the fits may be different. Thus the difference between the two results should be treated as a measure of the associated systematic error.

### *5.2. Vector mesons*

The vector decay constant is defined in terms of the matrix element of the local vector current $V_i$ between the rho and vacuum states

$$\langle 0 | V_i^{loc.} | \rho \rangle = \frac{\epsilon_i M_V^2}{Z_\psi^2 Z_V f_V}. \tag{5.2}$$

In our calculation $V_i^{local}$ is used to both create and annihilate the vector meson. The two methods used to extract $f_V^{-1}$ are described in Ref. [16] and we quote the mean value of the two values in Table 3 using three different choices for $Z_\psi^2 Z_V$.

*5.3. Analysis*

The data in Table 3 show that estimates for the decay constant have a very large dependence on the normalization of the currents. Results for $f_\rho^{-1}$ with the Lepage-Mackenzie scheme deviate by a large amount from experimental values as shown in Fig. 2 (case A in Table 3). Choosing a different value for the $q^2$ at which $\alpha_V$ is to be evaluated changes the numbers by $\lesssim 5\%$, which is insignificant compared to the difference at the heavier masses. The results using the non-perturbative value of $Z_V$ are in much better agreement (case C in Table 3 and Fig. 2). The non-perturbative values of $Z_V$, given in Table 4, are calculated using the ratios of 2-point correlators for rho mesons, for example $Z_V^{loc.}$ is defined as

$$Z_V^{loc.} = \frac{\langle V_i(t=0) V_i^{cons.}(\vec{p}=0,t)\rangle}{\langle V_i(t=0) V_i^{loc.}(\vec{p}=0,t)\rangle}, \tag{5.3}$$

with $Z_\psi = \sqrt{2\kappa^i}$ for $V_i^{cons.}$. The data for $Z_V^{ext.}$ is obtained using a similar ratio with $V_i^{loc.}$ replaced by $V_i^{ext.}$. Note that in the calculation of $f_\rho^{-1}$ using the above non-perturbative value of $Z_V$, the factors of $Z_\psi$ in $V_i^{loc.}$ cancel, so one cannot test a particular normalization scheme by this procedure.

In the case of the zero momentum correlator the vector current does not insert any momenta at the sink time-slice, so there should be no $O(a)$ corrections to the conserved current. This is why Eq. (5.3) is considered a good way to define $Z_V$. The data in Table 4 has been obtained using $Z_\psi = \sqrt{2\kappa^i}$ for $V_i^{loc.}$ and $V_i^{ext.}$ to allow comparison with previous published work. We find that $Z_V^{loc.}$ and $Z_V^{ext.}$ show very little dependence on the quark mass. Furthermore, the results are essentially unchanged when $Z_V$ is calculated from correlators at non-zero momentum. To convert to a different normalization, say Lepage-Mackenzie, requires multiplying the numbers in Table 4 by the ratio of the respective $Z_\psi$ factors, in which case $Z_V$ develop a dependence on the mass.

The same problem arises in the calculation of form-factors. The perturbative and non-perturbative normalizations give significantly different results and it is the improved perturbative normalization that seems to give consistent results for semi-leptonic form factors as we will discuss below. This dependence of the normalization on the nature of the initial and final state in the matrix element has been exposed and investigated by the Rome-Southhampton group [22]. In a recent study of semi-leptonic form factors they [23] have argued that this mismatch is a symptom of $O(a)$ effects and have presented data showing that the differences decrease on using the Sheikholeslami-Wohlert $O(a)$ improved action. Clearly this issue needs further study in view of the fact that the differences are so large.

## 6. Semileptonic form factors for $D^0 \to K^- e^+ \nu$

To orient the reader we first summarize the ways in which our method is technically different from those used in previous calculations.

1. We use smeared quark propagators to improve the signal while the previous calculations used point sources.
2. In our approach the interpolating operator for the final state meson sits at the source point ($t = 0$) of the original smeared light propagators. The source for the initial $D$ meson is at $t = 31$ and obtained by making a smeared pseudoscalar insertion in the light-to-heavy propagator $G^{ins}$. By using quark propagators calculated on doubled lattices we essentially eliminate wrap-around effects. With the two interpolating operators at a fixed separation, we can carry out the final contraction for the matrix element by placing the weak current at all intermediate times $0 < t < 31$. At each time slice $t$ we average the matrix element over all spatial points in a state of definite momentum transfer. One advantage of this method is that non-local lattice currents can be used to calculate the matrix elements. This approach is same as that used by the Rome-Southhampton group [5] but different from that used by Bernard *et al.* [3].
3. We calculate the matrix elements for three different transcriptions of the lattice current on the same set of lattices. Bernard *et al.* have only used $V^{local}$ while the Rome-Southhampton collaboration used only $V^{cons.}$.

Our calculation should be regarded more an exploration of methods rather than providing hard numbers that can be compared to experimental data because of the number of approximations made. We find that the data is of much better quality for the case where the final meson is annihilated by the pseudoscalar density $\overline{\psi}\gamma_5\psi$ as compared to the axial current $\overline{\psi}\gamma_4\gamma_5\psi$. For this reason we only present results for the pseudoscalar case.

In figs. 3-5 we show data for the ratios $\mathcal{R}_{LS}$ and $\mathcal{R}_{SS}$ for the three lattice transcriptions of the vector current. We have multiplied the $\mathcal{R}_{SS}$ data by $10^5$ and $\mathcal{R}_{LS}$ by $-1$ to put them on the same plot. The plots show that the signal is very similar with all three currents. When making fits to the data we find that $\mathcal{R}_{SS}$ has a better plateau and fits can be made using the full covariance matrix. The same is not true of $\mathcal{R}_{LS}$ and these results have been obtained without including the correlations.

To extract a composite quantity like the form-factor we have combined results from separate fits for each jackknife sample as our data sample is too small to make a simultaneous fit to all the necessary correlators. For example, to calculate $f_+$ we first make separate fits to the ratio $\mathcal{R}_{SS}$ for $V_4$ and $V_i$ and to the 2-point correlators $C^{ll}_{SS}$ and $C^{hl}_{SS}$ needed for the decay amplitudes occuring in Eqs. (4.6). A second estimate is similarly obtained by using Eq. (4.5). The difference between the two is a measure of the statistical uncertainty in the data. A third estimate is constructed by averaging the $\mathcal{R}_{LS}$ and $\mathcal{R}_{SS}$ results for each jackknife sample.

The results for all three estimates are given in Tables 5-7 using the normalizations defined in Eqs. (3.12), (3.14) and (3.13). There is a $\sim 20\%$ difference between the results obtained from $\mathcal{R}_{SS}$ and $\mathcal{R}_{LS}$ for the case of momentum transfer $\vec{p} = (\pi/8, 0, 0)$. This is a manifestation of the fact that the signal in non-zero momentum correlators falls very rapidly with $\vec{p}$ and this presents the largest source of error in the calculation of form-factors. We regard the third estimate (average of $LS$ and $SS$) as our best estimate, however, the reader should be aware that there is no deep reason for this.

We find that all three lattice transcriptions of the vector current give consistent results. The results with $V^{ext}$ are about 10% higher. This difference is of the same magnitude as that between the perturbative and non-perturbative value for $8\kappa_c$ and would be removed if we use the perturbative value for it in Eq. (3.14). It is this consistency which suggests that it is the improved normalizations for the currents that work well for the 3-point matrix elements.

The results for matrix elements in the forward direction are, within errors, 1.0 for all three currents. For non-zero momentum transfer our result for $f_+$ is consistent with the phenomenological estimate $f_+ \approx 0.75$, however, we have used too few values of quark mass and momentum transfer to determine whether pole-dominance is an accurate description of the variation of $f_+$ with respect to $Q^2$. We also find that $f_+ \approx f_0$ while the errors in $f_-$ are too large to make useful deductions.

If we express results using the normalization that seems to work for $f_\rho$, *i.e.* the improved $Z_\psi$ with the non-perturbative value for $Z_V$, then the results are no longer as consistent. For example the results for $f_+$ in Table 7 for $\kappa_l = 0.154$ get changed to 0.37(7), 0.42(11) and 0.60(9) for $V_\mu^{loc.}$, $V_\mu^{ext.}$ and $V_\mu^{cons.}$ respectively. The difference in results with the two schemes is substantial and indicates that the $O(a)$ effects with Wilson fermions are large and that the improved perturbative normalizations may not be very reliable for all matrix elements at $\beta = 6.0$.

## 7. Comparison with previous results

There are two previous calculations against which we can make direct comparisons as these use almost the same lattice parameters. The group of Bernard *et al.* [3] have measured the form-factors on $24^3 \times 40$ lattices at the same values of $\beta$ and $\kappa$. They used only the local vector current, and adopted yet a different normalization. Converting their result to our "improved" normalization gives $f_0(\vec{p}=0) = 0.85(10)$ at $\kappa = 0.154$ to be compared with our value of 0.91(9). Similarly the Rome-Southampton group [6] [7] have measured the form-factors on $20 \times 10^2 \times 40$ lattices at the same value of $\beta$ and similar $\kappa$. They use the "conserved" vector current with the same normalization for the vector current that we use. Interpolating their results to $\kappa = 0.154$ and noting that their momentum ($\vec{q} = 2\pi/10$) (and hence $Q^2$) is slightly different we get $f_+ = 0.67(5)$ and $f_0 = 0.65(4)$, to be compared with our results 0.60(9) and 0.62(10) respectively at ($\vec{q} = 2\pi/8$).

Thus we find that all three calculations give consistent results. The central remaining issue is to understand how to reduce the large uncertainty in the normalization of the currents.

## 8. Conclusions

We have given explicit details of how to self-consistently derive renormalization factors for fermionic operators to $O(\alpha_s)$. Our analysis reproduces the Lepage-Mackenzie mean-field results for local operators. We also show how to extend the analysis to non-local operators and to the Sheikholeslami-Wohlert improved action.

The results for decay constants and form-factors show large differences depending on the choice of normalization used for the currents. No one scheme seems to give reliable results for all the different matrix elements. These differences grow rapidly with the quark mass. Further work, possibly at weaker coupling and/or using $O(a)$ improved actions, is required to resolve this issue.

We find an acceptable signal in the 3-point correlators needed to calculate semi-leptonic form-factors using quark propagators with smeared sources. We show that differences in previous results is a consequence of different renormalization constants used for the vector current. Using the improved normalizations, we find that all three transcriptions of the vector current give consistent results. The biggest limitation in the calculation of form-factors is the rapidly decreasing signal in the non-zero correlators. Based on present data we believe that major improvements in results for the form-factors are possible by a high statistics study on larger lattices, say $32^3 \times 64$.

## Acknowledgements

We acknowledge the tremendous support received in the form of a DOE Grand Challenge award at NERSC at Livermore. A significant fraction of the calculation was done at the Pittsburgh Supercomputer Center and we are grateful to R. Roskies for his support. We thank G. Martinelli, S. Sharpe and A. Ukawa for useful discussions.

| | $\kappa_1 = 0.155$<br>$\kappa_2 = 0.155$ | $\kappa_1 = 0.154$<br>$\kappa_2 = 0.154$ | $\kappa_1 = 0.155$<br>$\kappa_2 = 0.135$ | $\kappa_1 = 0.154$<br>$\kappa_2 = 0.135$ | $\kappa_1 = 0.135$<br>$\kappa_2 = 0.135$ |
|---|---|---|---|---|---|
| $V^{loc}(imp.)$ | 0.231 | 0.235 | 0.270 | 0.272 | 0.316 |
| $V^{loc}(naive)$ | 0.218 | 0.217 | 0.204 | 0.203 | 0.190 |
| *Ratio* | 1.058 | 1.084 | 1.326 | 1.342 | 1.661 |
| $V^{ext}(imp.)$ | 0.280 | 0.286 | 0.328 | 0.331 | 0.384 |
| $V^{ext}(naive)$ | 0.266 | 0.265 | 0.249 | 0.248 | 0.232 |
| *Ratio* | 1.052 | 1.079 | 1.319 | 1.335 | 1.653 |
| $V^{cons}(imp.)$ | 0.310 | 0.308 | 0.289 | 0.288 | 0.270 |
| $V^{cons}(naive)$ | 0.310 | 0.308 | 0.289 | 0.288 | 0.270 |
| *Ratio* | 1.000 | 1.000 | 1.000 | 1.000 | 1.000 |
| $A^{loc}(imp.)$ | 0.249 | 0.253 | 0.291 | 0.294 | 0.340 |
| $A^{loc}(naive)$ | 0.240 | 0.238 | 0.224 | 0.223 | 0.209 |
| *Ratio* | 1.039 | 1.065 | 1.302 | 1.318 | 1.631 |

Table 1: The normalization constants for the three lattice transcriptions of the vector current and for the axial current used in this study for the 5 different combinations of quark masses. We give the improved and "naive" normalizations along with the ratio of the two for ease of comparison.

| | $\kappa_1 = 0.155$ $\kappa_2 = 0.155$ | $\kappa_1 = 0.154$ $\kappa_2 = 0.154$ | $\kappa_1 = 0.155$ $\kappa_2 = 0.135$ | $\kappa_1 = 0.154$ $\kappa_2 = 0.135$ | $\kappa_1 = 0.135$ $\kappa_2 = 0.135$ |
|---|---|---|---|---|---|
| $\pi^{SS}(0,0,0)$ | 0.299(9) | 0.362(6) | 0.813(5) | 0.833(5) | 1.216(4) |
| $\pi^{LS}(0,0,0)$ | 0.300(8) | 0.365(7) | 0.812(4) | 0.833(4) | 1.217(3) |
| $\pi_2^{SS}(0,0,0)$ | 0.297(16) | 0.369(10) | 0.815(8) | 0.836(7) | 1.217(4) |
| $\pi_2^{LS}(0,0,0)$ | 0.303(13) | 0.370(7) | 0.812(6) | 0.832(6) | 1.216(4) |
| $\pi^{SS}(1,0,0)$ | 0.5(1) | 0.49(3) | 0.89(1) | 0.91(1) | 1.265(7) |
| $\pi^{LS}(1,0,0)$ | 0.5(1) | 0.51(3) | 0.89(1) | 0.91(1) | 1.267(4) |
| $\pi_2^{SS}(1,0,0)$ | 0.5(1) | 0.52(3) | 0.90(1) | 0.92(1) | 1.268(7) |
| $\pi_2^{LS}(1,0,0)$ | 0.5(1) | 0.51(3) | 0.90(1) | 0.92(1) | 1.267(5) |
| $\rho^{SS}(0,0,0)$ | 0.407(16) | 0.459(14) | 0.842(8) | 0.862(7) | 1.229(4) |
| $\rho^{LS}(0,0,0)$ | 0.411(10) | 0.460(8) | 0.840(6) | 0.862(6) | 1.230(3) |
| $\rho^{SS}(1,0,0)$ | 0.6(1) | 0.60(3) | 0.93(2) | 0.94(2) | 1.278(8) |
| $\rho^{LS}(1,0,0)$ | 0.6(1) | 0.62(2) | 0.93(2) | 0.94(1) | 1.280(5) |
| $a_0^{SS}(0,0,0)$ | 0.7(1) | 0.7(1) | 1.08(5) | 1.07(4) | 1.42(3) |
| $a_0^{LS}(0,0,0)$ | 0.7(1) | 0.7(1) | 1.11(4) | 1.10(3) | 1.44(2) |
| $a_0^{SS}(1,0,0)$ | 0.7(1) | 0.8(1) | 1.2(1) | 1.1(1) | 1.47(3) |
| $a_0^{LS}(1,0,0)$ | 0.8(1) | 0.8(1) | 1.2(1) | 1.2(1) | 1.49(3) |

Table 2: Meson energies extracted from 2-point correlators. LS and SS denote local-smeared and smeared-smeared corrletors respectively. Two types of correlator were used for the to extract the pseudoscalar mass: (a) $\langle PP \rangle$ denoted $\pi$; and (b) $\langle A_4 A_4 \rangle$ denoted $\pi_2$. The energy is given for the particle at rest (the mass) and for the case of one unit of momentum. Quoted errors are statistical.

| | $\kappa_1 = 0.155$ $\kappa_2 = 0.155$ | $\kappa_1 = 0.154$ $\kappa_2 = 0.154$ | $\kappa_1 = 0.155$ $\kappa_2 = 0.135$ | $\kappa_1 = 0.154$ $\kappa_2 = 0.135$ | $\kappa_1 = 0.135$ $\kappa_2 = 0.135$ |
|---|---|---|---|---|---|
| (A) $f_\pi(0,0,0)$ | 0.076(5) | 0.089(5) | 0.125(4) | 0.131(4) | 0.206(4) |
| (A) $f_\pi(1,0,0)$ | 0.084(6) | 0.090(5) | 0.126(5) | 0.133(5) | 0.209(6) |
| (A) $f_\rho^{-1}(0,0,0)$ | 0.34(2) | 0.327(17) | 0.195(6) | 0.200(6) | 0.201(4) |
| (B) $f_\pi(0,0,0)$ | 0.073(7) | 0.084(6) | 0.098(4) | 0.101(3) | 0.126(4) |
| (B) $f_\pi(1,0,0)$ | 0.080(7) | 0.085(5) | 0.098(5) | 0.101(4) | 0.129(5) |
| (B) $f_\rho^{-1}(0,0,0)$ | 0.33(2) | 0.31(1) | 0.148(5) | 0.150(5) | 0.121(4) |
| (C) $f_\pi(0,0,0)$ | 0.082(7) | 0.093(6) | 0.107(4) | 0.111(3) | 0.141(5) |
| (C) $f_\pi(1,0,0)$ | 0.090(7) | 0.094(5) | 0.108(5) | 0.112(4) | 0.143(5) |
| (C) $f_\rho^{-1}(0,0,0)$ | 0.27(2) | 0.25(1) | 0.119(4) | 0.121(4) | 0.098(4) |

Table 3: The meson decay constants $f_P$ and $f_V$ extracted from $LS$ and $SS$ 2-point correlators. Quoted errors are statistical. The three cases differ in the normalization used for the currents. Case A uses the improved normalizations defined in this paper; case B uses $\psi^i_{cont} = \sqrt{2\kappa^i}\ \psi^i_L$ and perturbative values of $Z_A$ and $Z_V$ with boosted $g^2$; case C uses $\psi^i_{cont} = \sqrt{2\kappa^i}\ \psi^i_L$, $Z_A = 0.86$ and the non-perturbative estimate $Z_V = 0.57$.

| | $\kappa_1 = 0.155$<br>$\kappa_2 = 0.155$ | $\kappa_1 = 0.154$<br>$\kappa_2 = 0.154$ | $\kappa_1 = 0.155$<br>$\kappa_2 = 0.135$ | $\kappa_1 = 0.154$<br>$\kappa_2 = 0.135$ | $\kappa_1 = 0.135$<br>$\kappa_2 = 0.135$ |
|---|---|---|---|---|---|
| $Z_V^{\mathrm{loc.}}(\vec{p} = 0)$ | 0.572(5) | 0.571(3) | 0.565(2) | 0.565(3) | 0.551(2) |
| $Z_V^{\mathrm{loc.}}(\vec{p} = \parallel)$ | 0.55(2) | 0.57(1) | 0.544(8) | 0.545(7) | 0.536(5) |
| $Z_V^{\mathrm{loc.}}(\vec{p} = \perp)$ | 0.58(2) | 0.56(1) | 0.562(6) | 0.563(5) | 0.548(3) |
| $Z_V^{\mathrm{ext.}}(\vec{p} = 0)$ | 0.685(5) | 0.686(3) | 0.695(2) | 0.696(3) | 0.706(2) |
| $Z_V^{\mathrm{ext.}}(\vec{p} = \parallel)$ | 0.68(2) | 0.68(1) | 0.692(5) | 0.693(4) | 0.705(3) |
| $Z_V^{\mathrm{ext.}}(\vec{p} = \perp)$ | 0.69(2) | 0.70(2) | 0.690(7) | 0.693(6) | 0.704(3) |

Table 4: The renormalization constants $Z_V^{\mathrm{loc.}}$ and $Z_V^{\mathrm{ext.}}$ extracted from ratios of $LS$ 2-point correlators for the $\rho$ meson. These numbers have been obtained using $Z_\psi = \sqrt{2\kappa}$ normalization for all three vector currents. We give the results for zero momentum, and for one unit of momentum both parallel and perpendicular to the index of the current. Quoted errors are statistical.

| $\kappa = 0.154$ | | | | |
|---|---|---|---|---|
| Current | $f_+(Q^2 = -0.043)$ | $f_-(Q^2 = -0.043)$ | $f_0(Q^2 = -0.043)$ | $f_0(Q^2 = 0.222)$ |
| $V_\mu^{Local}$ | 0.56(13) | −0.33(23) | 0.58(14) | 0.91(12) |
| $V_\mu^{Ext.}$ | 0.60(12) | −0.34(24) | 0.62(14) | 0.98(11) |
| $V_\mu^{Cons.}$ | 0.55(11) | −0.20(16) | 0.56(11) | 0.88(13) |
| $\kappa = 0.155$ | | | | |
| Current | $f_+(Q^2 = -0.021)$ | $f_-(Q^2 = -0.021)$ | $f_0(Q^2 = -0.021)$ | $f_0(Q^2 = 0.268)$ |
| $V_\mu^{Local}$ | 0.59(19) | −0.54(30) | 0.61(20) | 0.91(9) |
| $V_\mu^{Ext.}$ | 0.60(23) | −0.43(31) | 0.61(24) | 1.00(10) |
| $V_\mu^{Cons.}$ | 0.55(20) | −0.22(25) | 0.56(21) | 0.87(9) |

Table 5: Form factors for the semi-leptonic decay $D \to K^- e^+ \nu$ using the ratio defined in Eq. (4.6). The results for $f_+$, $f_-$ and $f_0$ in columns 2-4 are for momentum transfer $\vec{q} = (\pi/8, 0, 0)$ while column 5 gives $f_0$ at $\vec{q} = 0$. Quoted errors are statistical.

| $\kappa = 0.154$ | | | | |
|---|---|---|---|---|
| Current | $f_+(Q^2 = -0.043)$ | $f_-(Q^2 = -0.043)$ | $f_0(Q^2 = -0.043)$ | $f_0(Q^2 = 0.222)$ |
| $V_\mu^{Local}$ | 0.67(12) | −0.56(31) | 0.73(15) | 0.91(10) |
| $V_\mu^{Ext.}$ | 0.76(18) | −0.47(30) | 0.81(18) | 1.02(15) |
| $V_\mu^{Cons.}$ | 0.65(10) | −0.25(20) | 0.68(11) | 0.89(9) |
| $\kappa = 0.155$ | | | | |
| Current | $f_+(Q^2 = -0.021)$ | $f_-(Q^2 = -0.021)$ | $f_0(Q^2 = -0.021)$ | $f_0(Q^2 = 0.268)$ |
| $V_\mu^{Local}$ | 0.71(23) | −0.76(42) | 0.78(24) | 1.01(13) |
| $V_\mu^{Ext.}$ | 0.73(25) | −0.61(41) | 0.78(25) | 1.10(14) |
| $V_\mu^{Cons.}$ | 0.67(22) | −0.33(33) | 0.70(21) | 0.98(12) |

Table 6: Form factors for the semi-leptonic decay $D \to K^- e^+ \nu$ using the ratio defined in Eq. (4.5). The rest is as in Table 5.

| $\kappa = 0.154$ | | | | |
|---|---|---|---|---|
| Current | $f_+(Q^2 = -0.043)$ | $f_-(Q^2 = -0.043)$ | $f_0(Q^2 = -0.043)$ | $f_0(Q^2 = 0.222)$ |
| $V_\mu^{Local}$ | 0.61(11) | −0.45(25) | 0.66(13) | 0.91(9) |
| $V_\mu^{Ext.}$ | 0.68(13) | −0.41(24) | 0.72(14) | 1.00(12) |
| $V_\mu^{Cons.}$ | 0.60(9) | −0.23(17) | 0.62(10) | 0.88(9) |
| $\kappa = 0.155$ | | | | |
| Current | $f_+(Q^2 = -0.021)$ | $f_-(Q^2 = -0.021)$ | $f_0(Q^2 = -0.021)$ | $f_0(Q^2 = 0.268)$ |
| $V_\mu^{Local}$ | 0.65(20) | −0.65(36) | 0.69(21) | 0.96(10) |
| $V_\mu^{Ext.}$ | 0.66(24) | −0.52(36) | 0.70(24) | 1.05(11) |
| $V_\mu^{Cons.}$ | 0.61(21) | −0.28(29) | 0.63(20) | 0.93(10) |

Table 7: Form factors for the semi-leptonic decay $D \to K^- e^+ \nu$ using the average of $LS$ and $SS$ data. The rest is as in Table 5.

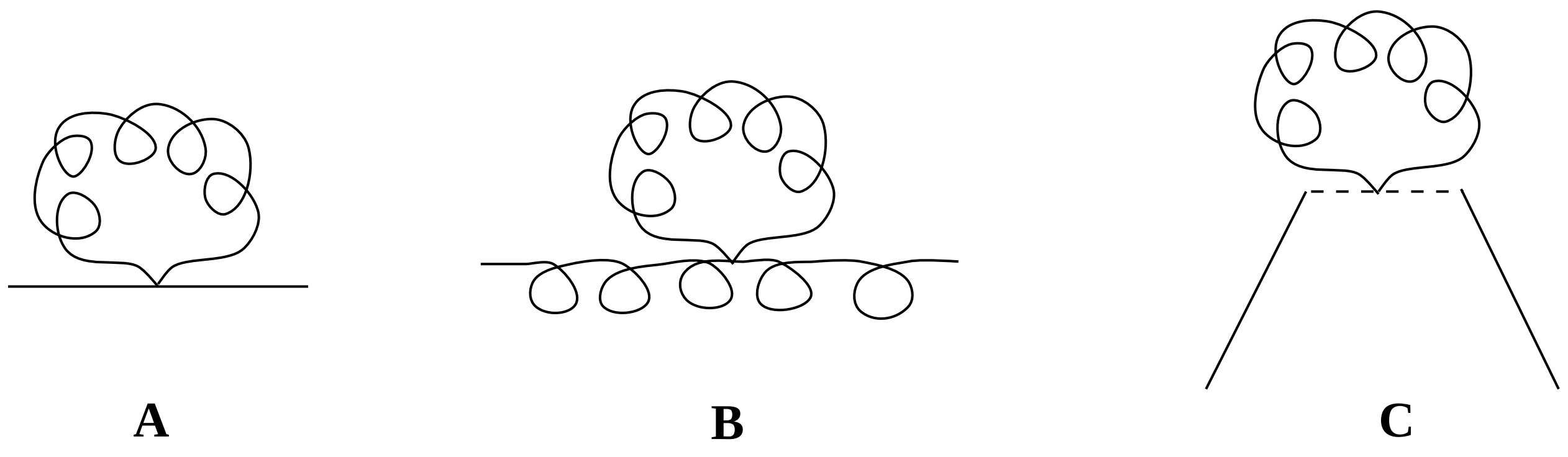

Fig. 1: The tadpole contributions to (a) fermion line, (b) gluon line, and (c) operator.

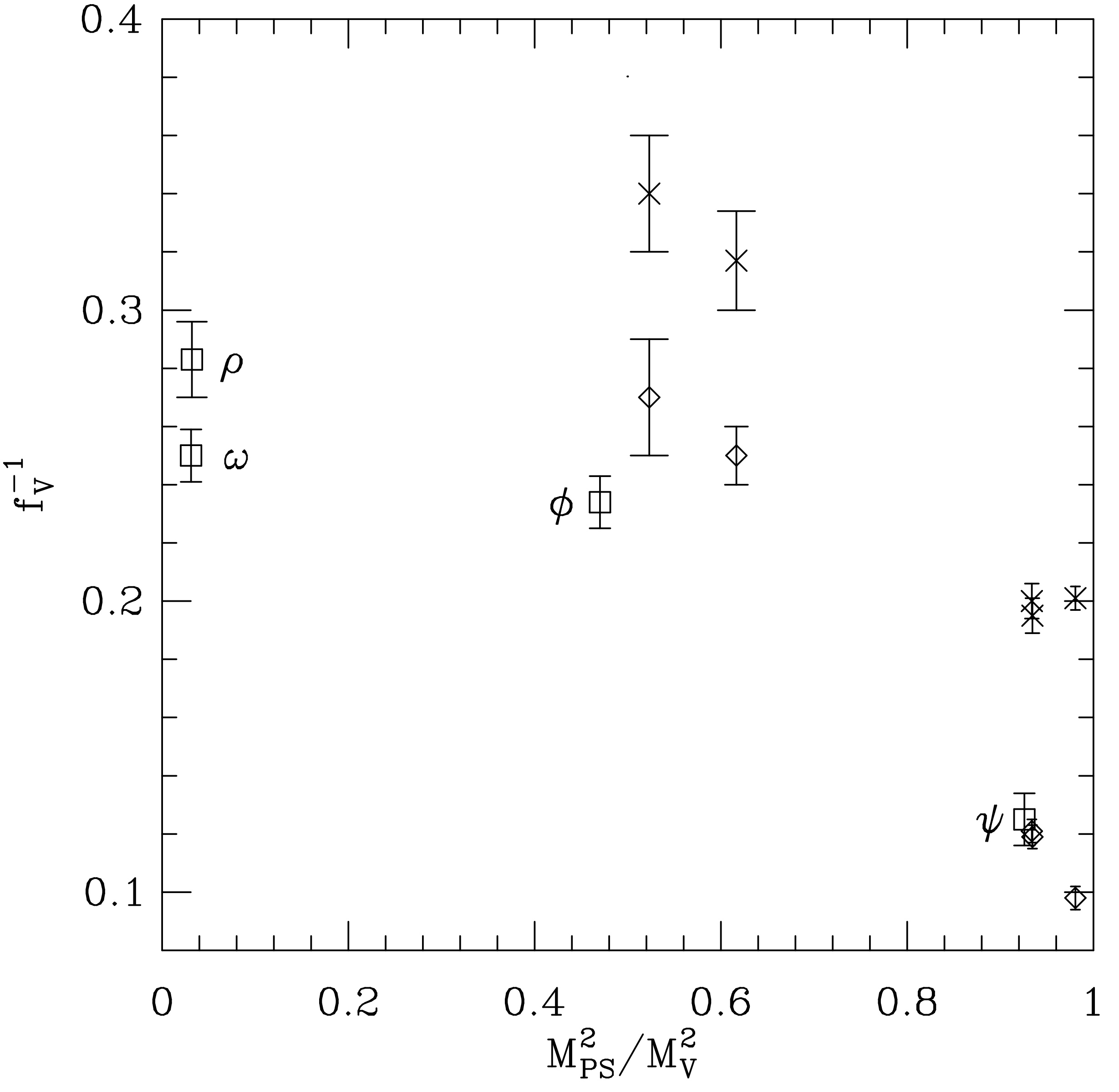

Fig. 2: The vector decay constant $f_\rho^{-1}$ as a function of $m_\pi^2/m_\rho^2$. The experimental numbers are labeled by the symbol $\Box$. The data given as case A (C) in Table 3 are marked by the symbol $\times$ ($\diamond$).

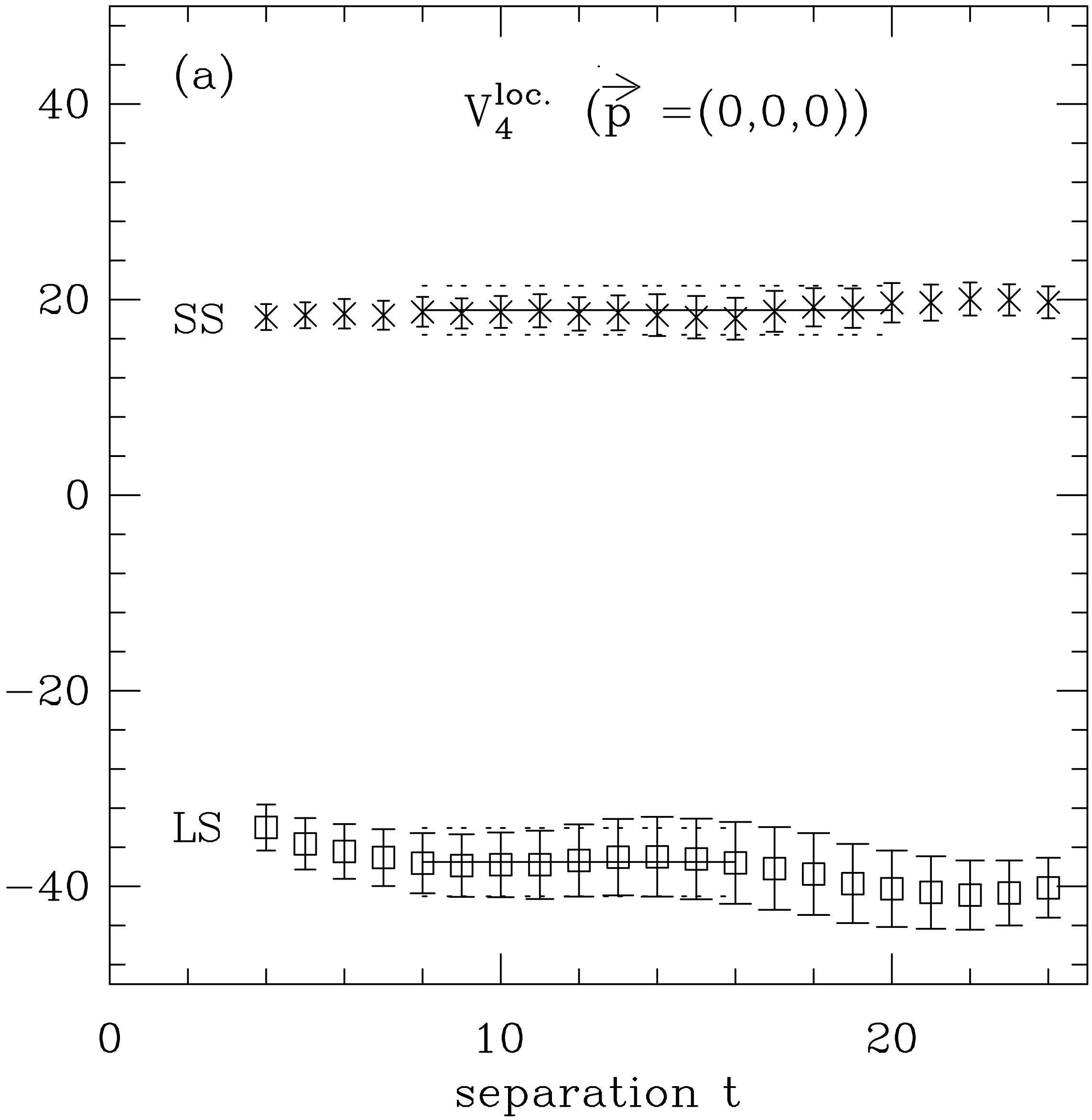

Fig. 3: The data and fit to $\mathcal{R}_{LS}$ and $\mathcal{R}_{SS}$ versus the seperation from the kaon's annihilation time-slice using the local vector current for (a) $V_4(\vec{p} = (0, 0, 0)$, (b) $V_i(\vec{p} = (\pi/8, 0, 0))$, and (c) $V_4(\vec{p} = (\pi/8, 0, 0)$.

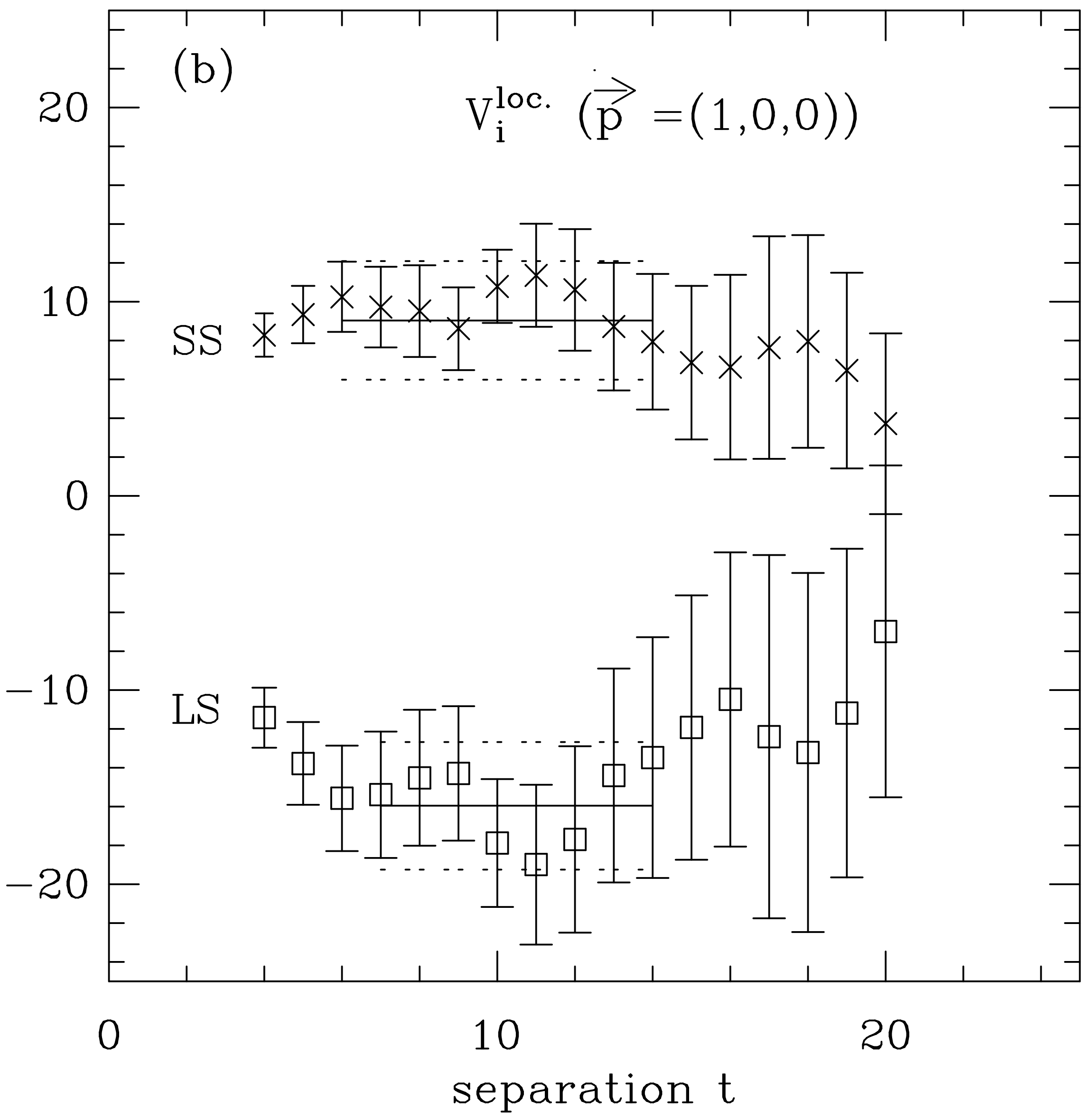

Fig. 3: The data and fit to $\mathcal{R}_{LS}$ and $\mathcal{R}_{SS}$ versus the seperation from the kaon's annihilation time-slice using the local vector current for (a) $V_4(\vec{p} = (0, 0, 0)$, (b) $V_i(\vec{p} = (\pi/8, 0, 0))$, and (c) $V_4(\vec{p} = (\pi/8, 0, 0)$.

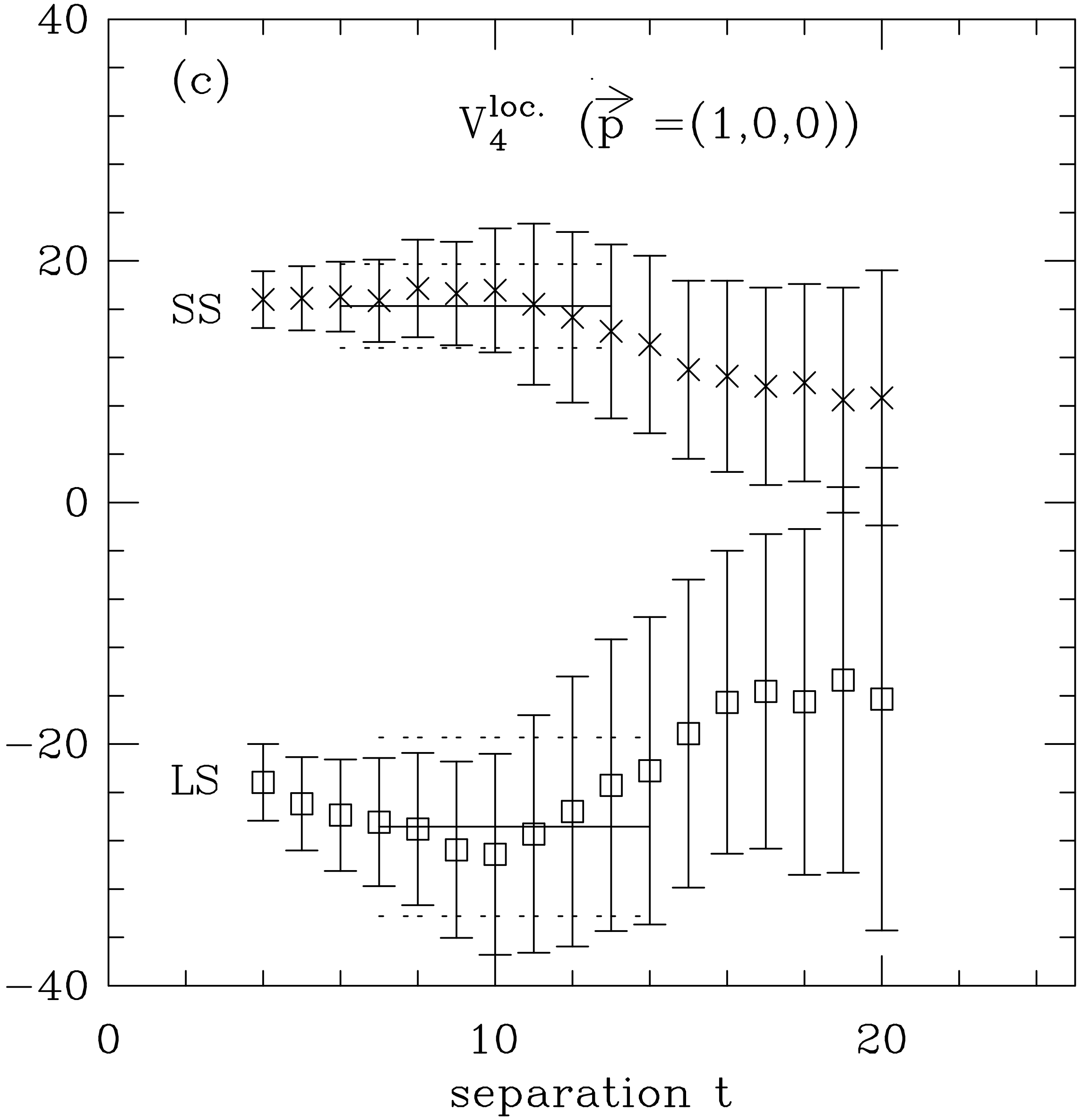

Fig. 3: The data and fit to $\mathcal{R}_{LS}$ and $\mathcal{R}_{SS}$ versus the seperation from the kaon's annihilation time-slice using the local vector current for (a) $V_4(\vec{p} = (0,0,0)$, (b) $V_i(\vec{p} = (\pi/8,0,0))$, and (c) $V_4(\vec{p} = (\pi/8,0,0)$.

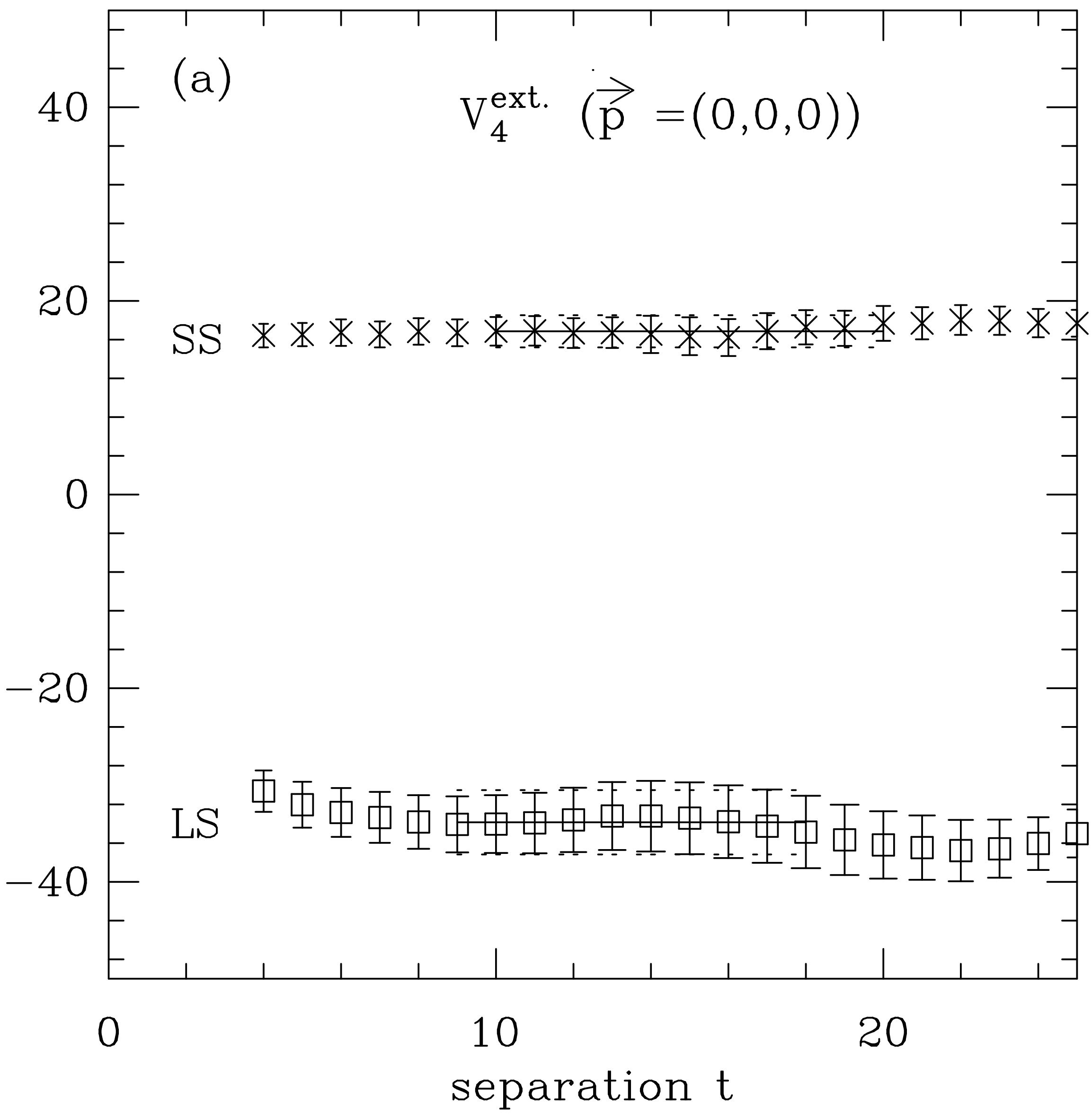

Fig. 4: The data and fit to $\mathcal{R}_{LS}$ and $\mathcal{R}_{SS}$ versus the seperation from the kaon's annihilation time-slice using the extended vector current for (a) $V_4(\vec{p} = (0,0,0)$, (b) $V_i(\vec{p} = (\pi/8,0,0))$, and (c) $V_4(\vec{p} = (\pi/8,0,0)$.

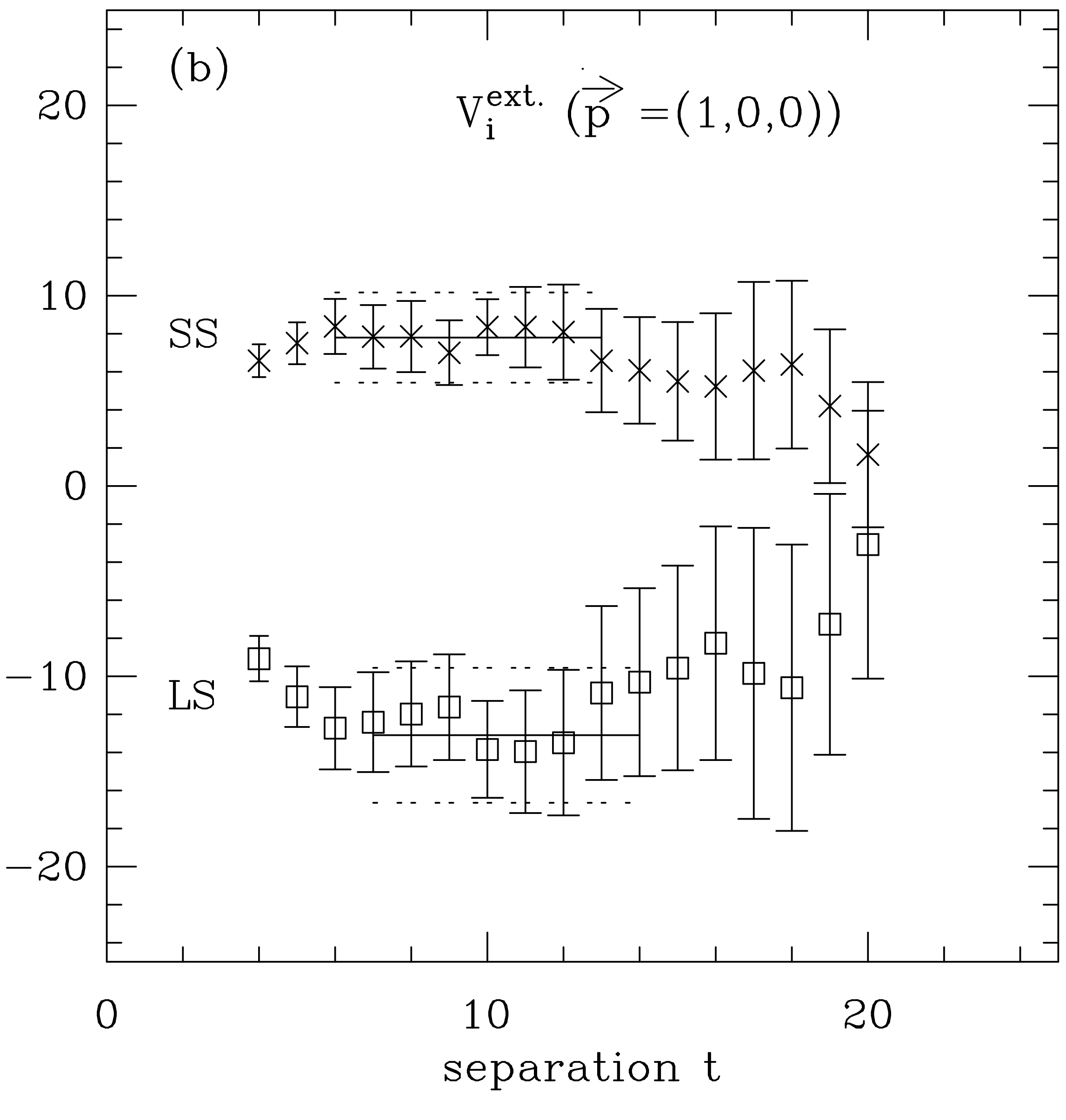

Fig. 4: The data and fit to $\mathcal{R}_{LS}$ and $\mathcal{R}_{SS}$ versus the seperation from the kaon's annihilation time-slice using the extended vector current for (a) $V_4(\vec{p} = (0,0,0)$, (b) $V_i(\vec{p} = (\pi/8,0,0))$, and (c) $V_4(\vec{p} = (\pi/8,0,0)$.

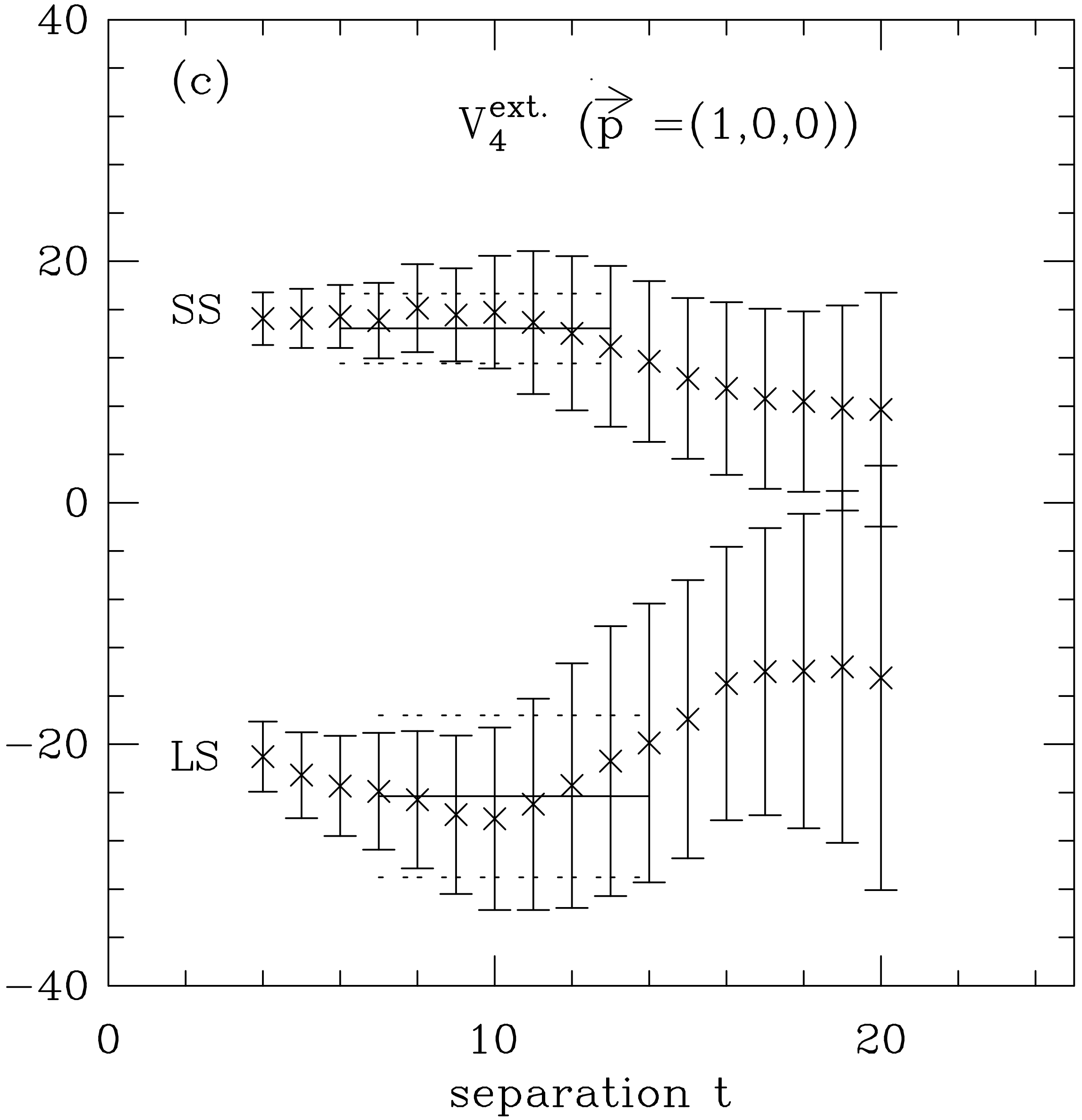

Fig. 4: The data and fit to $\mathcal{R}_{LS}$ and $\mathcal{R}_{SS}$ versus the seperation from the kaon's annihilation time-slice using the extended vector current for (a) $V_4(\vec{p} = (0,0,0)$, (b) $V_i(\vec{p} = (\pi/8,0,0))$, and (c) $V_4(\vec{p} = (\pi/8,0,0)$.

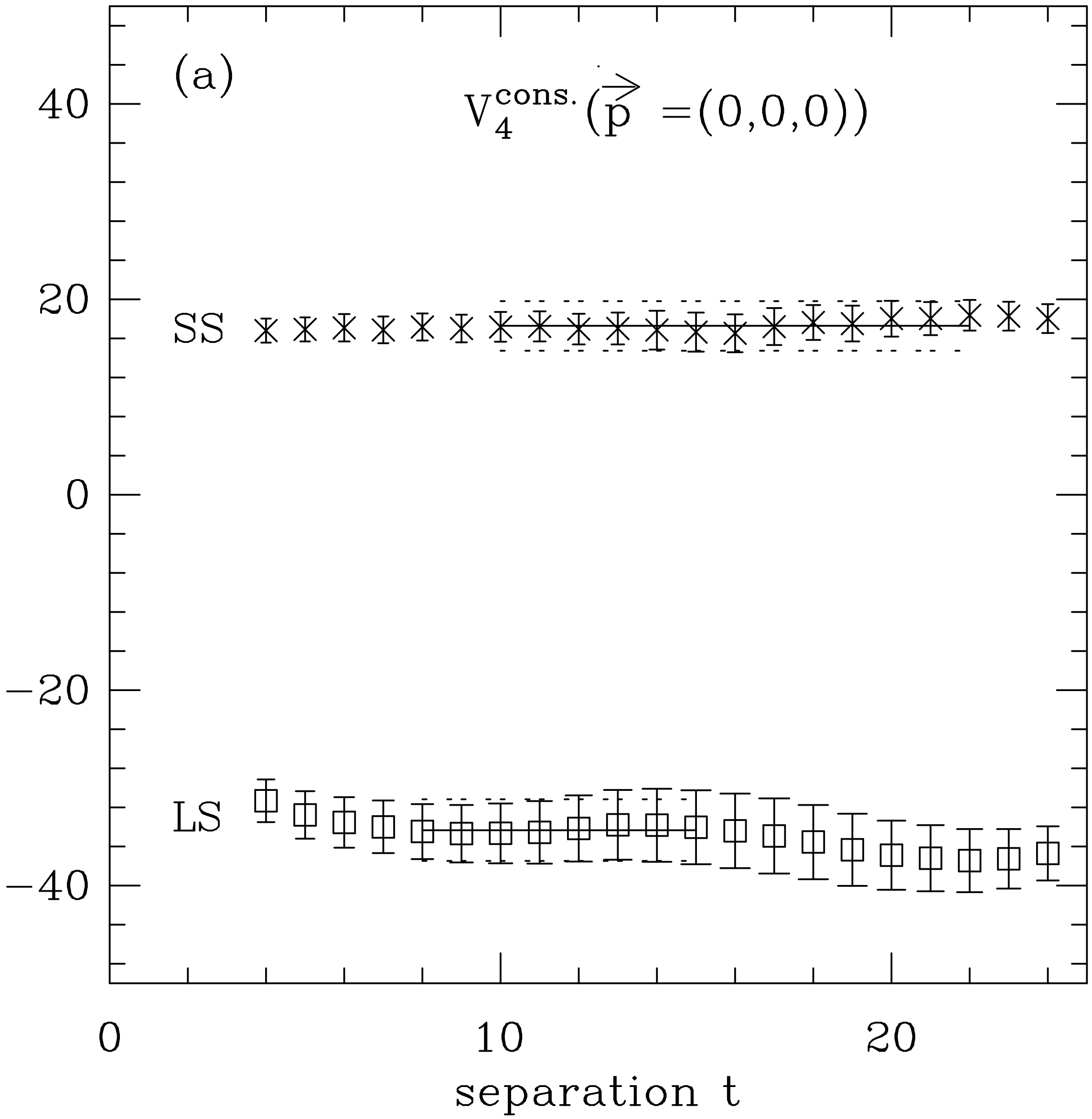

Fig. 5: The data and fit to $\mathcal{R}_{LS}$ and $\mathcal{R}_{SS}$ versus the seperation from the kaon's annihilation time-slice using the conserved vector current for (a) $V_4(\vec{p}=(0,0,0)$, (b) $V_i(\vec{p}=(\pi/8,0,0))$, and (c) $V_4(\vec{p}=(\pi/8,0,0)$.

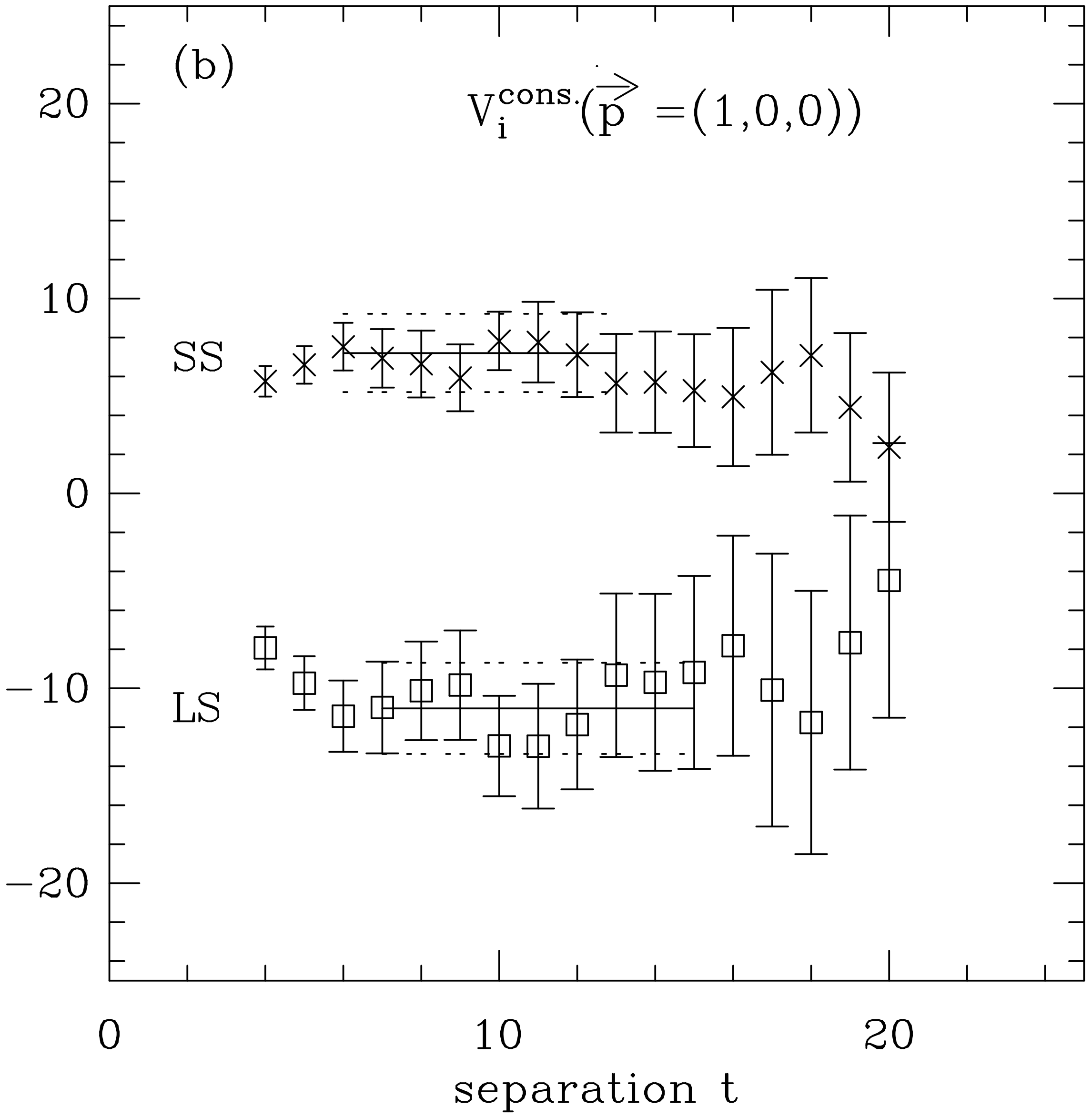

Fig. 5: The data and fit to $\mathcal{R}_{LS}$ and $\mathcal{R}_{SS}$ versus the seperation from the kaon's annihilation time-slice using the conserved vector current for (a) $V_4(\vec{p}=(0,0,0)$, (b) $V_i(\vec{p}=(\pi/8,0,0))$, and (c) $V_4(\vec{p}=(\pi/8,0,0)$.

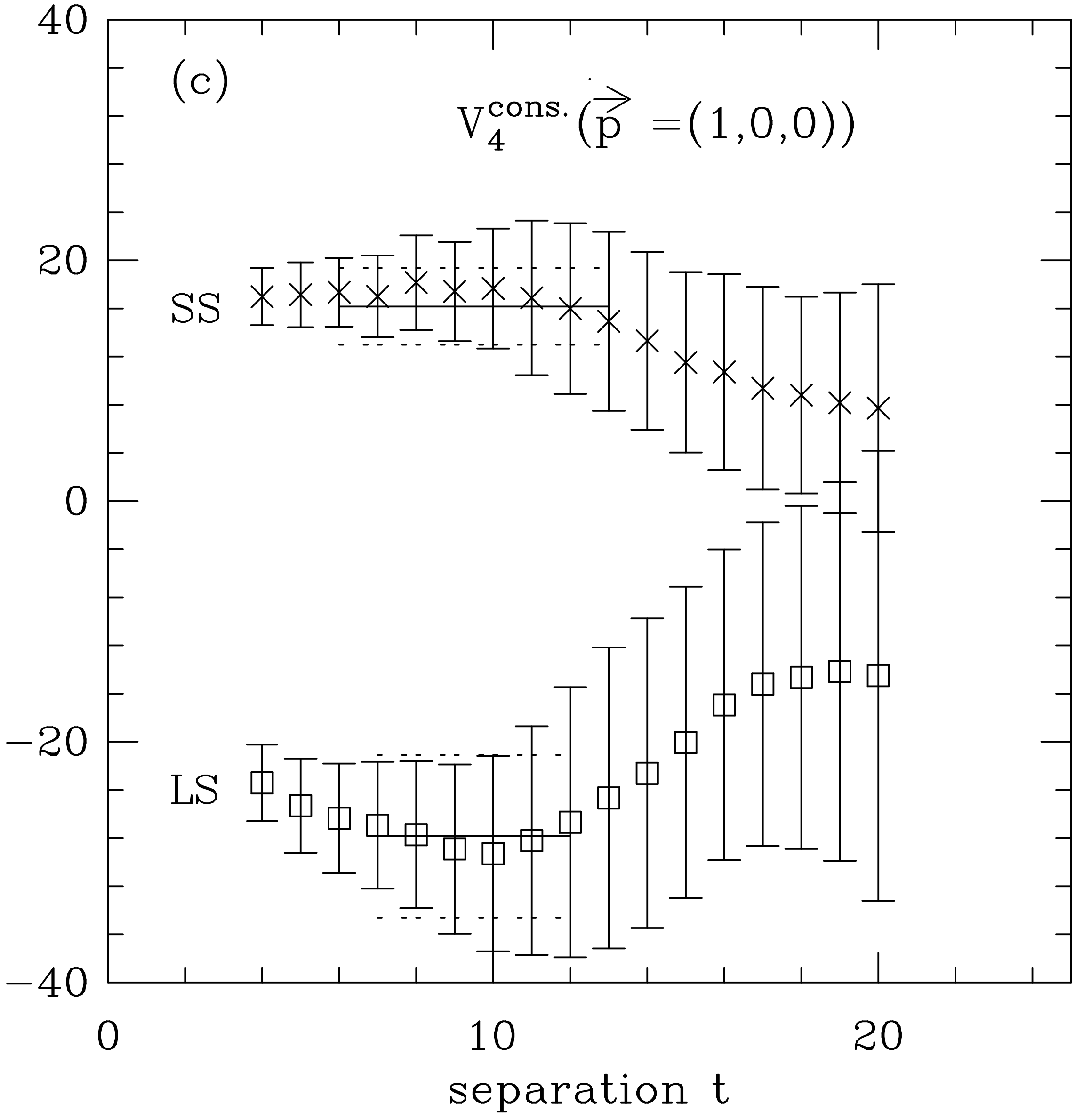

Fig. 5: The data and fit to $\mathcal{R}_{LS}$ and $\mathcal{R}_{SS}$ versus the seperation from the kaon's annihilation time-slice using the conserved vector current for (a) $V_4(\vec{p} = (0,0,0)$, (b) $V_i(\vec{p} = (\pi/8,0,0))$, and (c) $V_4(\vec{p} = (\pi/8,0,0)$.